# physica status solidi

www.pss-journals.com

*reprint*

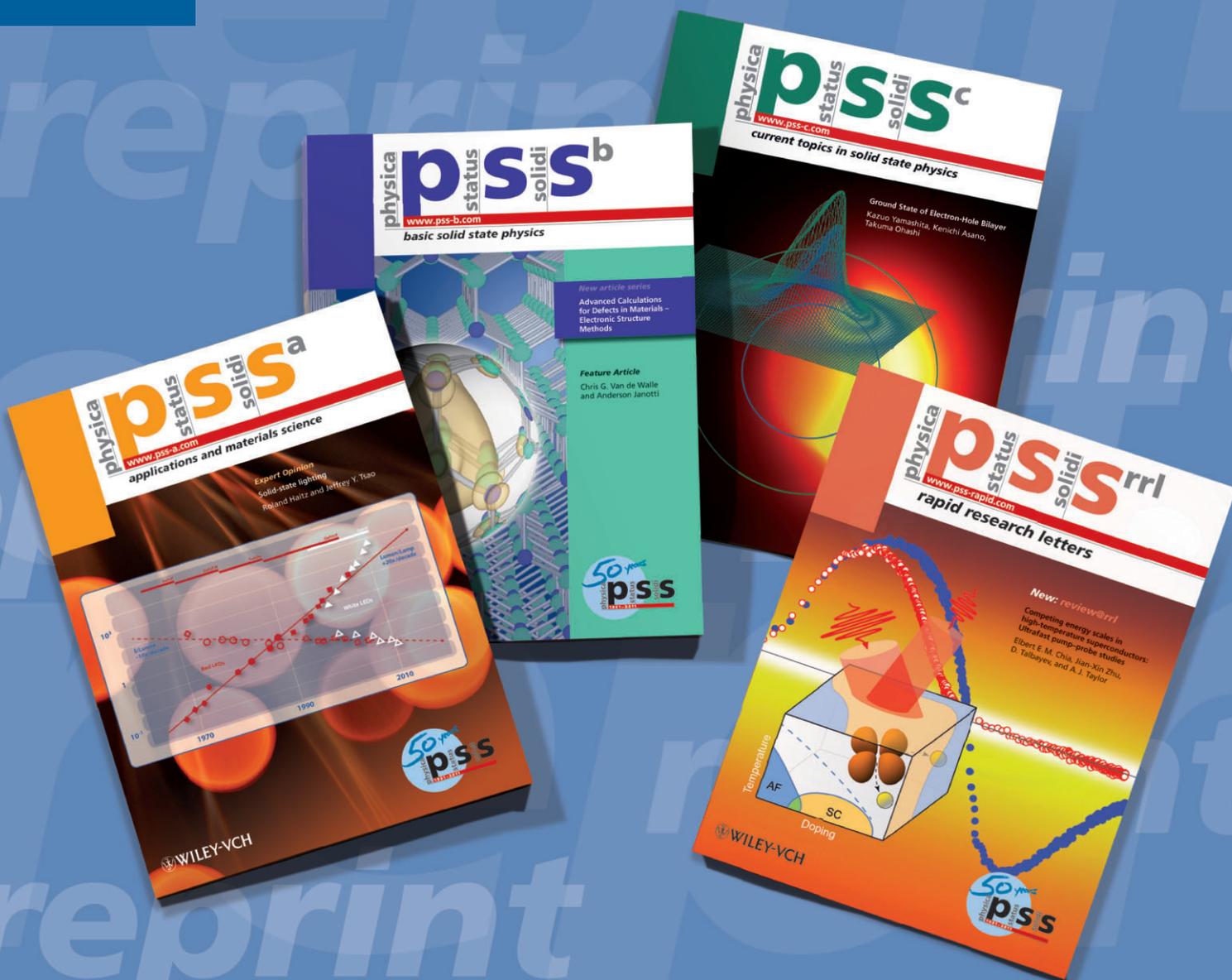

WILEY-VCH

# Kinetic properties of the two-dimensional conducting system formed by CrSi$_2$ nanocrystallites in plane (111) of silicon

V. V. Andrievskii[1], Yu. F. Komnik[1], I. B. Berkutov[*,1,2], I. G. Mirzoiev[1], N. G. Galkin[3], and D. L. Goroshko[3]

[1] B. Verkin Institute for Low Temperature Physics and Engineering of the National Academy of Sciences of Ukraine, 47 Lenin Ave., Kharkov 61103, Ukraine
[2] Department of Physics and Astronomy, Wayne State University, 666 West Hancock Ave., Detroit, MI 48201, USA
[3] Institute of Automation and Control Processes of the Far Eastern Branch of Russian Academy of Sciences, 5 Radio Street, 690041 Vladivostok, Russia

The behaviors of resistance, magnetoresistance (up to 5 T), and Hall electromotive force (EMF) with varying temperature (10–300 K) and measuring current (10 μA–10 mA) are studied for the Si sample with CrSi$_2$ nanocrystallites (NC) in the plane (111). The conduction in such heterostructure proceeds in the plane with the NC and is the conduction of a two-dimensional system of charge carriers that shows some unusual effects. The temperature variation of resistivity may be treated as the result of the effect of thermal activation but in this case it is characterized by a low activation energy different in value in different temperature ranges. This suggests that the mechanism of conduction is more complex. It is found that the conduction is determined by the effect of temperature variation not only on carrier concentration but also on its mobility. Magnetoresistivity is also of different shape in different temperature ranges. All the above features are treated in terms of the proposed model of electron hopping through the conduction band (or hole hopping through the valence band). A peculiar effect of giant reduction in resistivity with increasing the measuring current has been revealed. Discussed are some possible factors responsible for this effect.

© 2013 WILEY-VCH Verlag GmbH & Co. KGaA, Weinheim

**1 Introduction** The silicides are one of the promising materials for microelectronics. Semiconductor silicides can be used as materials of transistors and metal silicides as stable contacts to these transistors. During deposition of some metals on silicon substrates, formation of both metal and semiconductor silicides takes place [1] so the transistor and contact platform can be grown during the same technological operation.

Transition metal silicides have many attractive properties for microelectronic applications due to their good conductivity, high melting point, excellent oxidation resistance [1–4]. Also owing to the good compatibility with silicon substrates, some metal silicides show high potential in various silicon-based optoelectronics and spintronics applications [5, 6]. Additionally, possibility of the epitaxial growth of transition metal silicides on silicon surface has motivated attempts to create silicon/silicide heterostructures [7, 8].

One of the most interesting and well-investigated compounds is chromium disilicide, which has the smallest lattice mismatch with monocrystalline silicon in comparison with other transition metal silicides [1, 8]. Chromium disilicide CrSi$_2$, being a narrow-band semiconductor with energy gap $E_g = 0.32$ eV, is one of the most extensively studied semiconducting silicides [1] and is widely used in creation of optoelectronic IR elements and thermoelectronic elements. It is also employed as a highly sensitive anisotropic radiation detector with a low noise level [9, 10]. Taking into consideration the temperature stability of chromium silicides and their different electrical properties depending on the composition and thickness, this system is very interesting for research.

The formation of metal and semiconductor silicides on silicon substrates has found wide interest since it may fulfill requirements in silicon planar technology [11–13]. The Cr–SiO and Cr–Ge thin films has been prepared and intensively studied in amorphous [14–16] and nanodisperse [17] state. Since chromium silicides may be grown epitaxially, they were of interest not only for new device concepts, but they are of fundamental interest o understand the special transport properties of very thin epitaxial layers [18].

Progress in semiconductor technology has made it possible to create various semiconductor single crystals in which impurity atoms are located within one crystallographic plane (the so-called δ-layers). This results in the appearance of two-dimensional conductivity, similar to that in heterojunctions or inversion layers. But, δ-layers were little used in microelectronics mainly because of the low mobility of the carriers subjected to frequent scattering at the ionized atoms in the δ-layer [19]. Meantime a structure in which one crystallographic plane of a high-energy-gap semiconducting crystal contains nanodimensional crystallites, rather then individual atoms, can exhibit new interesting and practically significant properties. Semiconductor materials created on the basis of buried nanocrystallites (NC), including semiconductor silicide quantum dots in silicon matrix, can possess new optical and electric properties, which are important for construction of new device structures. The growth of chromium disilicide NC (10–40 nm) as a high density δ-layer of CrSi$_2$ NCs/Si heterojunctions yields novel electrical, thermoelectrical, and optical properties. The quantum size effect will be observed under greatly decreasing CrSi$_2$ crystallite sizes (4–6 nm [20]) which results in quantization of energy levels and increase of effective band gap value. The last can alter the fundamental transition type: indirect–direct for CrSi$_2$ [1]. In the case that quantum-size CrSi$_2$ crystallites are densely($10^{10}$–$10^{11}$ cm$^{-2}$) buried in the silicon crystal lattice one could expect hopping conduction between the NC. This conduction depends on the properties and defect structure of the CrSi$_2$ NCs/Si interfaces which can considerably alter transport, optical, electrical, photoelectric, and thermoelectric characteristics. These can be used to design new types of apparatus structures [21]. Optical, photoelectric and thermoelectric properties of such systems are studied adequately [7, 22] in contrast to their transport properties.

In this work, the kinetic properties of the system of CrSi$_2$ NC grown in the plane (111) of silicon were studied in a wide range of magnetic fields (up to 5 T) with varied temperature (10–300 K) and measuring current (10 μA–10 mA). The investigation of temperature and electric field dependencies of the resistance of the system under study shows some unusual effects. A possible mechanism of the charge carrier transport responsible for the observed peculiarities is discussed.

**2 Sample description** The experiment was concerned with a Si crystal where the CrS$_2$ NC were localized in the plane (111). The detailed description of the sample preparation and characterization given in Ref. [21]. The sample was cut from Si(111) substrates of p-type conductivity and 1 Ω cm resistivity. The silicon cleaning procedure was as follows: annealing at 700 °C for 4–5 h, cooling for 12 h, flushes at 1250 °C (five times in 1 s). Surface purity was controlled by AES and by LEED methods. After the preparation procedure, the silicon surface showed the bright Si(111) 7 × 7 LEED pattern. Chromium was deposited on hot (500 °C) silicon substrate in the thickness range of 0.01–0.18 nm (0.125–2.25 Ml). Silicon layers were deposited atop chromium disilicide islands from a rectangle-shaped silicon sublimation source heated with direct current. Chromium and silicon deposition rates were controlled by quartz sensor. Chromium was deposited at ∼0.017 nm min$^{-1}$ (0.021 Ml min$^{-1}$). Silicon deposition rate was ∼4 nm min$^{-1}$ (20.8 Ml min$^{-1}$ on Si(111) substrate). On the basis of electron microscopy, the sample displayed two types of CrSi$_2$ NC: small ones of size ∼3 nm and large ones of size ∼20–40 nm. The NC were 2–4 nm in height. The average distance between small NC was ∼20 nm and their surface density was ≈2.5 × 10$^{11}$ cm$^{-2}$. The surface density of large NC was ≈3 × 10$^9$ cm$^{-2}$. The conduction in such a heterosystem is realized in the plane of the NC, that is, it is the conduction of a two-dimensional electron (hole) system.

The paper concerns the measurements of temperature dependences of resistance, magnetoresistance, and Hall electromotive force (EMF) in the temperature range 10–300 K. For dc electric measurements the area of the sample with CrSi$_2$ NC was a Hall bar having width ∼1.5 mm and length 9 mm. The magnetic field strength up to 5 T was produced with a superconducting solenoid with an automatic field scan.

**3 Temperature dependence of resistivity** The sample studied exhibited semiconductive-type variations in resistivity with temperature. The dependences $\rho_{xx}(T)$ obtained at measuring currents of 3, 5, 10, and 50 μA [1] are shown in Fig. 1. As is evident at temperatures below ∼20 K, the resistivity decreases slightly with increasing temperature, but reduces considerably with increasing current. At temperatures above 20 K, the curves obtained at currents of 3, 5 and 10 μA are completely coincident. This allows the value 10 μA to be taken as a "standard" for the measurement current at $T > 20$ K. [2]

Figure 2 shows the dependences of $\ln \rho_{xx}$ on $1/T$ which permit us to estimate the fulfillment of the Arrhenius equation of temperature variation of resistivity with the activation process [23],

$$\rho(T) = \rho_0 \exp\left(\frac{E_i}{k_\mathrm{B} T}\right), \tag{1}$$

in different temperature ranges.

The whole temperature range (10–300 K) studied was arbitrarily separated into several temperature intervals in which, as will be shown below, the behaviors of resistance, magnetoresistance, and Hall EMF were significantly different.

---

[1] The resistivity scale is logarithmic.
[2] The use of lower values may result in a considerable increase of the experimental error.

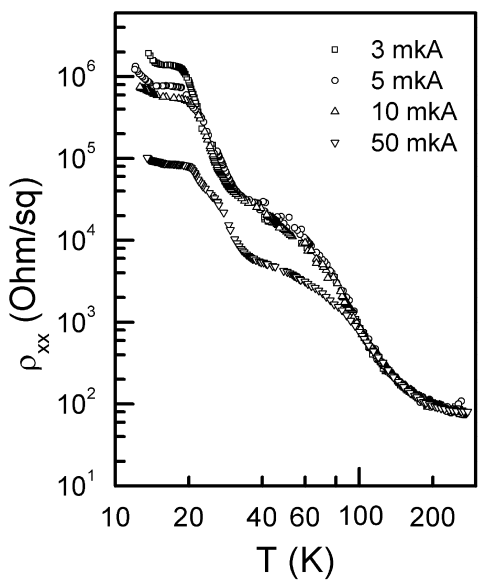

**Figure 1** Temperature dependence of resistivity $\rho_{xx}(T)$ for different values of currents.

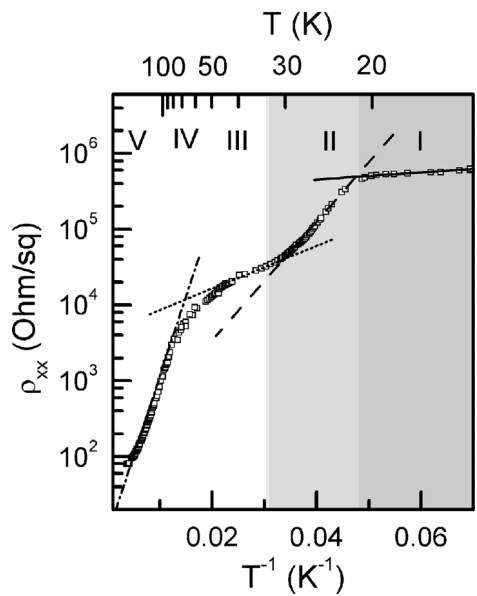

**Figure 2** Sample resistivity (logarithmic scale) as a function of $1/T$ for measuring current of $10\,\mu A$. The straight lines are the experimental data described by the Arrhenius relation (Eq. (1)) for different temperature ranges: the solid line is for region I, the dashed line is for region II, the dotted and dash-dotted lines are for regions III and V, respectively.

In region I (below 20 K), energy $E_1$ in Eq. (1) (solid line in Fig. 2) is very low $\sim 0.9$ meV (at $10\,\mu A$). This may suggest that the conductivity in this region is realized through the well-known variable-range hopping mechanism. The electrons (or holes) localized near the $CrSi_2$ NC, make jumps between vacant states, and this process, in the first approximation, is dependent on current value. The hopping mechanism takes place in the forbidden band of Si.

In region II (from 20 to $\sim 40$ K), the activation energy is $E_2 = 6.88$ meV (dashed line in Fig. 2), but in region III (from 40 to $\sim 70$ K) it reduced down to $E_3 = 2.86$ meV (dotted line in Fig. 2). Next is a transition interim region IV (from 70 to $\sim 90$ K) and for the temperatures between 100 K and $\sim 200$ K (region V) the activation energy is $E_5 = 42$ meV (dash-dotted line in Fig. 2).

It should be mentioned that a possible existence of dissolved Cr atoms in silicon would result in an activation energy of $\sim 0.3$ eV [24] and the activation processes would contribute to the temperature variation of resistivity only at higher temperatures. An increase in carrier density with temperature is principally determined by the processes of carrier escape from the $CrSi_2$ NC.

The temperature dependencies of resistance of amorphous [14, 15] and nanodisperse [17] Cr-SiO thin films been studied in the wide temperature range (50 mK–300 K). In these study the resistivity obeys the Mott's law [25]: $\rho(T) = \rho_0 \exp(T_0/T)^\alpha$ with $\alpha$ changing from 1/4 (at $T \approx 100$ K) to 1 with decreasing temperature for amorphous films and with $\alpha = 1/2$ for nanodisperse films. Here $T_0 = \beta/k\nu(\mu)a^3$, $\nu(\mu)$ is the density of states at the Fermi level, $a = \hbar/(2m\varepsilon_1)^{1/2}$ and $\beta = 21.2$ is a numerical coefficient [23]. These dependencies can be explained by variable-range hopping that is phonon-assisted tunneling between localized states with energies near the Fermi energy. For a two-dimensional system, Mott's law variable range hopping is $\rho(T) = \rho_0 \exp(T_0^*/T)^{1/3}$, where $T_0^* = \beta^*/kg^*(\mu)a^2$, $g^*(\mu)$ is the two-dimensional density of states at the Fermi level, and $\beta^* = 13.8$ [26, 27]. This behavior was observed in Ref. [28] for temperature dependencies of resistance of Si single crystals with $\delta(Sb)$ layers at sufficiently low temperatures ($<10$ K). But at higher temperatures variation of the resistance was described well by Eq. (1) with $E \approx 43$ meV which is similar to that obtained in present study for region V.

The temperature ranges with a very low activation energy, where the Arrhenius law (regions II and III) is fulfilled, demonstrate that the process of carrier transport is determined not only by thermal activation but it is a more complex one. Indeed, the surprising thing is that the activation energy $E_3$ in higher temperature region III is much lower than the activation energy $E_2$ in low-temperature region II. The activation process intensity is to be kept constant or even to increase with increasing temperature.

The above features of the temperature variations of resistivity in regions II and III is considered below on the basis of the following concept: The principal discrepancy between the mechanism of conductivity for the object studied and the mechanism of impurity conductivity of a semiconductor consists in that the appearance of carriers yielding the transport properties is stimulated by the emission of carriers from the $CrSi_2$ NC into silicon. Before passing into the silicon crystal, the carriers must penetrate the heteroboundary between the two semiconductors – chromium disilicide and silicon. This process is realized by tunneling through the surface barrier (Schottky barrier). The escape of electrons (or holes) from the NC causes the them to be charged. It is just the charge that determines the specific features of the system studied.

The CrSi$_2$ NC are deposited in a single crystallographic plane (111) of silicon. In this plane there occurs a deformation, of the Si energy spectrum in response to the charged NC. If the NC are positively charged, they form hollows (cavities) in the bottom of the conductivity band of silicon which are gradually occupied by the activated electrons. Similar hollows also form in the ceiling of the valence band. If the NC are negatively charged they form bulges at the valence band ceiling where holes are concentrated. Similar bulges form also in the bottom of the conduction band. The hollows and bulges are quantum wells for electrons and holes where the carriers occupy the quantum level. As an illustration, we consider below a version of electron conductivity with hollows, while a version of hole conductivity necessitates the change from hollows to bulges and from conduction band to valence band.

The electron travel in response to electric field is realized in electron motion through the conduction band by the tunneling effect (maybe in combination with a slight activation process) and it resembles a hopping conductivity. But unlike the well-known variable-range hopping between acceptors or donors in the forbidden band of a semiconductor [23] which is commonly characterized by a low mobility, the expected hopping motion of electrons occurs in the conduction band, providing a rather high mobility (see below). In real space, the electrons in quantum wells are localized nearby the NC, but they have an energy higher than the Fermi energy.

The existence of hollows (or bulges) in response to the charge on NC results in a slight decrease of the activation energy because of reducing the distance in the energy scale between the initial state and those in quantum wells. However, the existence of the charge on NC causes a more global consequences. It is possible that the joint charge on NC in the crystallographic plane (111) of silicon reduces the energy distance from the initial state to the conduction band bottom for electrons (or to the valence band ceiling for holes). This supposition explains the low activation energy obtained.

We suggest that in temperature region II the thermal activation results in the occupation of large bulges associated with large NC with a high charge. At a certain temperature this process is saturated and then there occurs an occupation of small bulges related to minor NC in temperature region III. This model explains the difference in activation energy values in the Arrhenius equation for region II and III.

To obtain data on temperature dependence of carrier density $n$ and mobility $\mu = e\tau/m^*$ ($e$ is the charge of electron, $\tau$ is the transport relaxation time, and $m^*$ is the effective mass of electron) the Hall EMF $U_{xy}$ was measured at different temperatures. It was found that in all cases the Hall EMF varied linearly with magnetic field, permitting to estimate the Hall constant values of the two-dimensional system under consideration. The Hall constant as well as carrier density $n$ and mobility $\mu$ were estimated in terms of the two-dimensional model. For one type of carriers in the conduction equation the values for conductivity $\sigma$ and $n$ are related to a unit of area, and the Hall constant equation is of the form $R_H = U_{xy}/IB$, where $I$ is the current and $B$ is the magnetic field strength.

It should be noted that for the object studied the Hall constant varies with temperature over wide limits: as the temperature is increased from 25 to 100 K, it decreases by an order of five (from $10^5$ to $10^0$ T$^{-1}$). As is evident from the measurements of Hall EMF, the conduction of the sample studied is p-type conductivity.

The quantities $n$ and $\mu$ were calculated by the standard equations for one type of carriers:

$$\sigma = ne\mu, \qquad (2)$$

$$R_H = \frac{1}{ne}, \qquad (3)$$

in different temperature ranges. The value of density is obtained by Eq. (3) and that of mobility by Eq. (2) or by the expression $\mu = \sigma R_H$.

The carrier density increases considerably with temperature in regions II and III (Fig. 3a). In region IV this rise slows down for an unknown reason and in region V it is kept constant at the same level. The Hall constant in regions IV and V ceases to decrease with increasing temperature for the reason that a second group of carriers with considerably different characteristics may appear. It seems likely that at these temperatures the conductivity of layers adjacent to the plane with NC becomes commensurable to that of the plane.

The carrier mobility in regions II and III decreases with increasing temperature (Fig. 3b) and in region IV it is kept constant and even increases in region V. This unnatural and inexplicable variation of $\mu$, as well as the behavior of density $n$ (Fig. 3a), suggests that it is quite possible that Eqs. (2) and (3) cannot be taken for regions IV and V; instead, the calculation scheme for a two-layer system should be applied. If we assume that in regions IV and V the conductivity is determined by two groups of carriers with essentially different characteristics, the open points in Fig. 3 are the averaged values for two groups of carriers.

The dramatic variations in carrier mobility with temperature in regions II and III motivates a correction to the estimation of activation energy in these regions. For these cases one should use the formula that gives a direct description of variations in concentration

$$n(T) = n_0 \exp\left(-\frac{\varepsilon_i}{k_B T}\right). \qquad (4)$$

The estimation by Eq. (4) gives the following values of $\varepsilon_i$: for region II $\varepsilon_2 = 17$ meV and for region III $\varepsilon_3 = 6.8$ meV. Elimination of temperature dependence of $\mu$ resulted in that the activation energy increased by the factor 2.4.

Of special note is the fact that the carrier mobility in region II is very high for such an inhomogeneous sample. This supports the suggestion made in the proposed model that the conduction in regions II and III is realized by moving electrons through the conduction band (holes through the valence band) although it is of a hopping behavior.

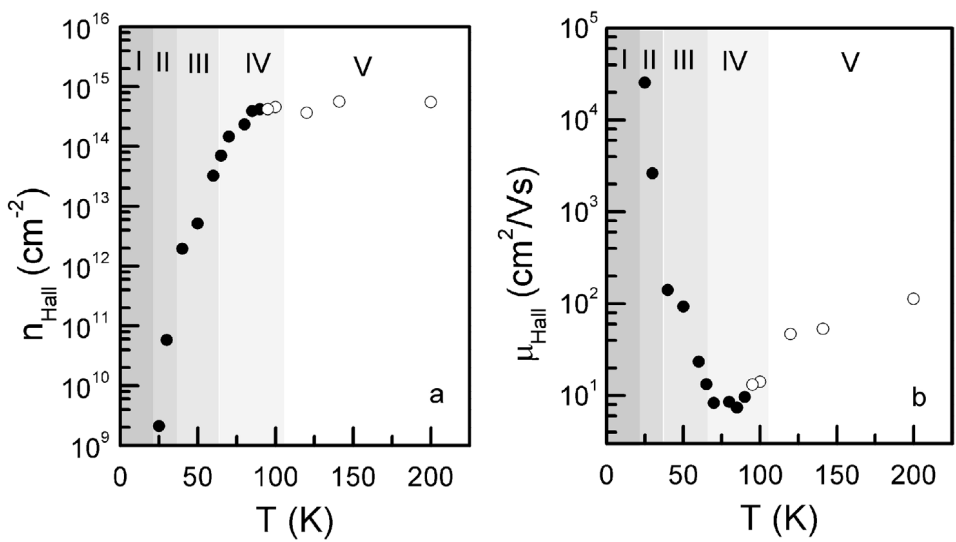

**Figure 3** Carrier density $n$ (a), and mobility $\mu$ (b) for different temperatures calculated by Eqs. (2) and (3).

**4 The effect of giant variations of resistance in response to current.** Figure 4 shows the profound decrease in the sample resistivity observed with increasing the measuring current. This effect is particularly dramatic at low temperatures and gradually diminishes with increasing temperature.

The data shown in Fig. 4 were used when plotting the dependence of conductivity $\sigma$ on applied voltage $U$ (Fig. 5) at temperatures above 40 K. Such dependences at lower tem-peratures were not constructed because of a weak $U$ to $I$ relationship. As is evident from the Fig. 5, an increase in conductivity at all the temperatures is defined by the current applied to the sample. From the initial portions of the curves up to $U \approx 1$ V this relation is of the form $\sigma \propto \exp(AU)$; the coefficient $A$ decreases slightly at temperatures above 90 K (in region V).

According to the relation (2) an increase in conductivity may be due to (i) a rise of carrier density or (ii) an increase in carrier mobility. The first option suggests that there exists a carrier emission from NC under the effect of electric field. The second option suggests the existence of the effect of electron overheating (an increase in carrier temperature as compared to that of the crystal lattice in response to a high electric field) at which the carrier mobility may increase.

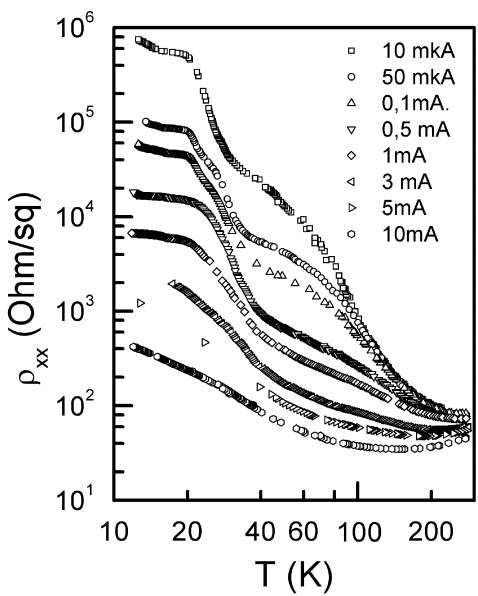

**Figure 4** Temperature dependences of sample resistivity for different values of current.

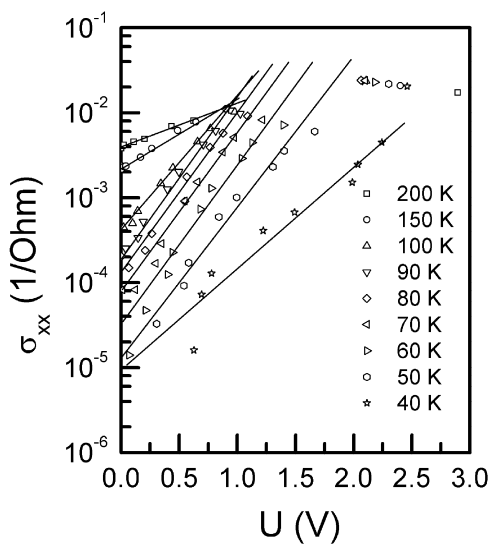

**Figure 5** Conductivity $\sigma_{xx}$ as a function of voltage $U$ for different temperatures.

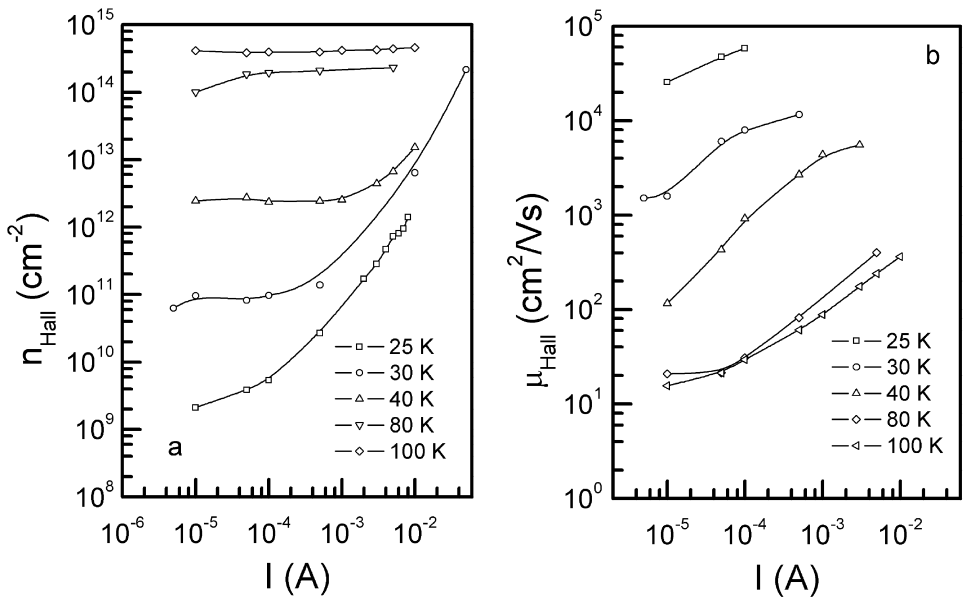

**Figure 6** Carrier density $n$ (a) and mobility $\mu$ (b) as a function of current for different temperatures (the solid lines are plotted to guide the eye).

In the model of carrier emission in response to electric field the above-mentioned relation between $\sigma$ and $U$ can be obtained if we suggest that the appearance of excess carriers is determined (like the Richardson–Dashman formula for thermoelectric emission [29]) by the relation

$$n_{\text{ex}} = \exp\left(-\frac{\varphi - eU}{k_{\text{B}} T}\right), \qquad (5)$$

where $\varphi$ is the work function. Lowering of the surface potential barrier by the $eU$ value increases the probability of electron escape by the factor $\exp(eU/k_{\text{B}} T)$.

To check the carrier emission model, the Hall EMF was measured at currents varied from 10 mkA to 10 mA at different temperatures. Using the data for Hall EMF, the values of carrier density were calculated by Eq. (3). The expected increase in $n$ with current was observed only at the lowest temperatures (25, 30, and 40 K; Fig. 6a). But for this temperature range one cannot convert one-to-one the current values to the applied voltage due to high sample resistance. Figure 6a supports qualitatively the supposition of carrier emission in response to the applied voltage.

Nevertheless, starting from the current dependences of carrier density, one can draw several conclusions. From the comparison of Figs. 4 and 6a it follows that the carrier emission in response to current cannot properly explain the effect of conductivity enhancement. For instance, if at 25 K a rise in carrier density with increasing current from 10 μA to 10 mA is almost commensurable with the rise in conductivity, then at $T = 40$ K the variation of current within the above range produces a fivefold increase in density while the conductivity increased by a factor of 250. The measurements of Hall EMF at higher temperatures (regions IV and V) have not revealed an increase in carrier density with current.

Whereas, these measurements demonstrate that the carrier mobility increases with increasing current (Fig. 6b). This supports the second option that explains the effect under consideration.

The significant role in this case, particularly at temperatures above 40 K, is likely to be played by the effect of electron overheating. In the object under study this effect is realized in a peculiar way. The effect of electric field comes normally to an increase in electron temperature. In the object under consideration a rise in electron energy occurs due to another reason. The conduction band bottom under the influence of an electric field is inclined along the sample. Therefore, the electrons, that move by the tunnel effect along the field from one quantum well to another, find themselves in a higher energy state with respect to the bottom of quantum well. In a time of transition through a certain portion of the path an averaged energy over the quantum well bottom rises, while the chemical potential for the whole carrier system is kept constant. The increase in carrier energy is not associated with the direct energy absorption from the electric field. For electrons, moving opposite to field direction, the transition from one quantum well to another requires expenditure of energy, that is, it is related to the activation process.

This peculiar process of carrier "heating" results supposedly in an increase of effective electron mobility and thus in a rise of conductivity.

**5 Magnetoresistivity** Variations in magnetoresistivity $(\rho(B) - \rho_0)/\rho_0$ of the sample with temperature are shown in Fig. 7. In region II the magnetoresistivity (MR) displays a linear dependence (Fig. 7a) and has a remarkably high value. This may be due to a low carrier density, activated in the quantum wells, and corresponds to the high mobility observed in this region (see Fig. 3b). With increasing temperature, the MR value reduces rapidly. In region III the low magnetoresistivity remains positive and linear (Fig. 7b). But beginning with 60 K, there appears a nonlinear portion in the low magnetic field range, and then for $T = 70–90$ K the magnetoresistivity becomes negative (Fig. 7c). This is because along with the positive MR there appears a negative component with saturation at high magnetic field. At temperatures above 90 K the magnetoresistivity is positive (Fig. 7d), low and exhibits a quadratic rise at high magnetic fields (higher than 4T).

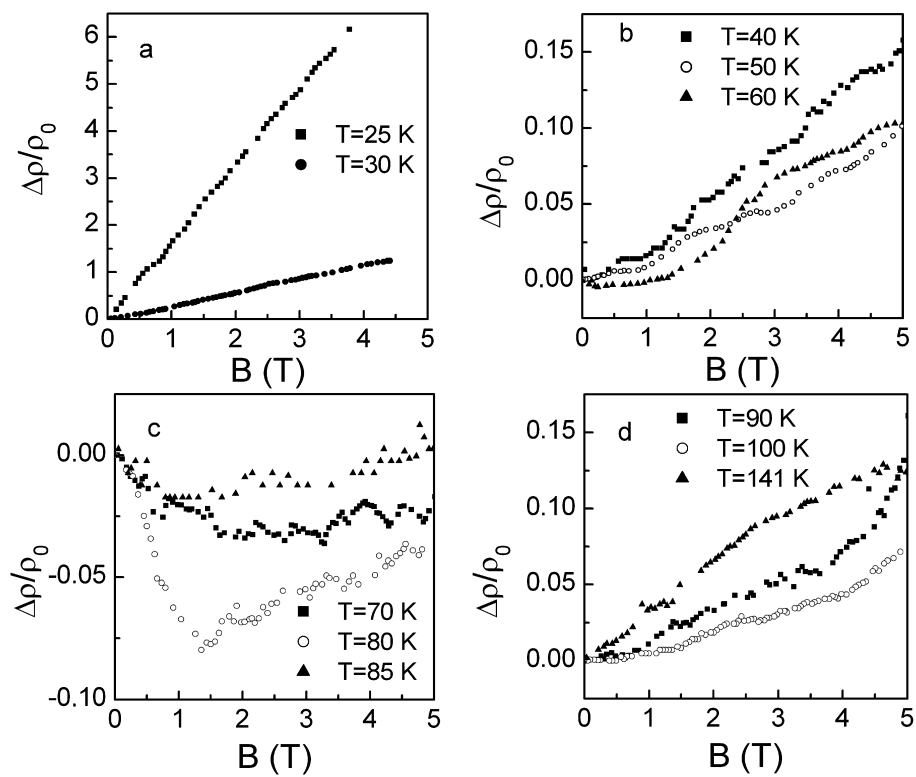

**Figure 7** Magnetoresistivity $(\rho(B) - \rho_0)/\rho_0$ as a function of magnetic field $B$ for different temperature intervals: a) region II, b) region III, c) region IV, d) region V.

We suggest that in interim region IV the quantum wells in the conduction band (or in the valence band) are overfilled by activated carriers, resulting in the appearance of delocalized carriers that are capable of providing a normal band conductivity over the quantum wells. This may be responsible for a certain increase in the averaged mobility in regions IV and V (Fig. 3b) as well as for the appearance of negative magnetoresistivity.

For two-dimensional (2D) electron systems a quasi-classical mechanism of resistance changes in magnetic field has been proposed in Ref. [30]. This mechanism gives a negative magnetoresistivity for a typical case. The authors of Ref. [30] draw attention to an essential difference in the behavior of electrons in magnetic fields between three- and two-dimensional cases. In a three-dimensional electron gas the cyclotron motion of electrons in a high magnetic field proceeds along the field, and the electrons inevitably encounter with an impurity and are scattered by it. In a two-dimensional case for the magnetic field normal to the 2D system, electron motion along the field is impossible. In this case there are electrons which move infinitely in a circular path about an impurity or, colliding with this impurity, form a rosette-type path. The electron motion around the impurity implies electron localization near the impurity and such electrons make no contribution to current. For a small scatterer concentration a negative magnetoresistivity should appear. Papers [31, 32] present model numerical calculations of magnetoresistivity for a 2D system of electrons scattering by randomly distributed impurities and proves the existence of negative MR in accordance with the predictions given in Ref. [30]. A more detailed analysis of the problem is presented in Ref. [33] that also demonstrates the appearance of negative MR saturating with increasing magnetic field. Thus, the negative magnetoresistivity in region IV can be accounted for by the appearance of free carriers in the conduction (or valence) band.

**6 Conclusions** An original concept is proposed to explain the specific features of the kinetic properties of the two-dimensional electron system formed by NC located in a single crystallographic plane of a semiconductive crystal. It is supposed that under the effect of nanocrystallite charges the conduction band bottom of the semiconductor (with electron conductivity) or the valence band ceiling of the semiconductor (with hole conductivity) are deformed, resulting in the formation of quantum wells ("hollows" or "bulges") at the conduction band bottom (or valence band ceiling). The carrier transport is realized by transitions between localized quantum states in the quantum wells. These transitions occur by tunneling, similarly to the well-known variable-range hopping conductivity in the forbidden band of a semiconductor with impurity atoms. But the "interhollow" ("interbulge") hopping conductivity is realized through the conduction (or valence) band and, as known, the band conductivity is characterized by a rather high mobility. This mechanism involving the processes of thermal activation of carriers makes it possible to explain the temperature dependence of resistance for the sample with CrSi$_2$ NC in the plane (111) of silicon as well as a variety of magnetic field-resistance relations.

One more specific feature of the object studied is the observed effect of giant decrease in resistance with increasing measuring current. Discussed are two possible causes for this effect: (1) emission of carriers from NC in response to electric field and (2) overheating of carriers when moving along the quantum wells resulting in a mobility increase.